\documentstyle[11pt,epsfig]{article}

\newcommand{\be}{\begin{equation}}
\newcommand{\ee}{\end{equation}}
\newcommand{\ba}{\begin{eqnarray}}
\newcommand{\ea}{\end{eqnarray}}
\begin{document}

\thispagestyle{empty}

\hfill  UCL-IPT-03-17

\begin{center}
{\Large \bf Relating Final State Interactions in $B\to D\pi$ and
$B\to DK$
\par} \vskip 1.5cm
{\large {\sc G.~Calder\'on$^1$, J.-M.~G\'erard$^2$,
J.~Pestieau$^2$ and J.~Weyers$^2$
}\\[1ex]
{\normalsize $^1$ \it Departamento de F{\'i}sica, Cinvestav del IPN\\
Apartado Postal 14-740, 07000, M\'exico D.F., M\'exico}\\[2ex]
{\normalsize $^2$ \it Institut de Physique Th\'eorique,
Universit\'e
catholique de Louvain\\
B-1348 Louvain-la-Neuve, Belguim}}

\par \vskip 1em

\end{center}

\begin{abstract}
A Regge model calculation relates the strong phase in $B \to DK$
to that in $B\to D\pi$. This provides a significant test of a
hadronic picture of final state interactions in $B$ decays.
\end{abstract}

\newpage

\par\noindent
{\Large \bf 1. Introduction}\vskip 0.5cm

Recently the CLEO Collaboration \cite{cleo} performed a detailed
amplitude analysis of the decays $B \to D\pi$: they determined the
absolute values of the isospin 1/2 and isospin 3/2 amplitudes as
well as their relative phase $e^{i\delta_{D\pi}}$ with the result

\begin{equation}
\cos \delta_{D\pi}  = 0.86 \pm 0.05
\end{equation}

\noindent indicating significant final state interaction effects.

On the other hand, the BELLE Collaboration \cite{belle} has
reported the first measurement of the decay $\overline B^0 \to D^0
\overline K^0$. With this new experimental information, a central
value analysis \cite{xing} of the decays $B\to DK$ suggests, once
again, important final state interaction effects although the data
remain compatible with a small relative phase $e^{i\delta_{DK}}$
between the isospin zero and isospin one amplitudes \cite{rosner}.

In our view these strong phases are genuine hadronic effects
which, we believe, cannot meaningfully be parametrized by pure
short-distance considerations. In Refs. \cite{gerard1,gerard2}, a
simple Regge model was proposed to calculate these strong phases.
The predictions of the model are in good agreement with the
experimental data for the decays $D \to \pi\pi$, $\pi K$ and
$K\overline K$. At present it is not yet possible to significantly
test the model in the corresponding $B$ decays.

In this note we extend this Regge model to the decays $B\to D\pi$
and $B\to DK$. From the short distance point of view these decays
differ radically from e.g. $B\to K\pi$ since there are no penguin
topology contributions. In a hadronic model for strong phases,
isospin symmetry is important while the underlying quark diagram
topology is basically irrelevant \cite{gerard3}. It is this idea
which we propose to test in the new class of $B$ decays.

We will first of all argue that a  sensible phenomenological model
is to identify $\delta_{D\pi}$ with $\delta_3-\delta_1$ namely the
difference in $s$-wave phase shifts of the $D\pi$ scattering
amplitudes in the isospin 3/2 and isospin 1/2 channels
respectively. A Regge model would then lead to a prediction of
$\delta_{D\pi}$ if the couplings of the Pomeron and $\rho$
trajectory to the $D\overline D$ channel were known. This would be
the case in an unrealistic SU(4) symmetry limit but it appears
more sensible to plot $\delta_{D\pi}$ in terms of a single
variable $x_{D\pi}$ which only depends on Regge parameters.

We then proceed to an analogous parametrization and calculation of
$\delta_{DK} = \delta_1-\delta_0$, i.e. the difference in phase
shifts in the isospin one and isospin zero amplitudes in $B\to
DK$. Once again $\delta_{DK}$ depends on a single parameter
$x_{DK}$.

The main  point of this note is to point out that $x_{DK}$ is
uniquely determined from $x_{D\pi}$. In other words a better
determination of $\delta_{D\pi}$ would lead to a precise
prediction of $\delta_{DK}$. The present data are certainly
compatible with this prediction which lies at the heart of a
hadronic approach to final state interactions.

\clearpage

\par\noindent
{\Large \bf 2. Hadronic final state interactions}\vspace{0.5cm}

The asymptotic states of QCD are hadrons, not quarks and gluons.
Isospin invariance is an excellent symmetry of the hadronic world,
hence the S-matrix for strong interactions  commutes with the
isospin generators.

The standard decomposition of $B\rightarrow D\pi$ decays in terms
of isospin amplitudes reads

\ba A(B^- \rightarrow D^0 \pi^-)\ &=&\ \sqrt{3}\
A_{\frac{3}{2}}\nonumber\\
A(\bar{B}^0 \rightarrow D^+ \pi^-)\ &=&\
+ \sqrt{\frac{2}{3}}\ A_{\frac{1}{2}} + \frac{1}{\sqrt{3}}\
A_{\frac{3}{2}} \nonumber\\
A(\bar{B}^0 \rightarrow D^0 \pi^0)\ &=&\ - \frac{1}{\sqrt{3}}\
A_{\frac{1}{2}} +\sqrt{\frac{2}{3}}\ A_{\frac{3}{2}} .\ea

Phenomenologically it makes good sense to view the isospin
amplitudes as being built up from a direct (weak) transition
followed by rescattering. This is embodied in the standard formula
\cite{suzuki}

\be A_I(B\rightarrow i)\ =\ \langle i,out|H^{(2)}_W|B\rangle_I \ =
\ \sum_{j} S^{\frac{1}{2}}_{ij}\ \bar{A}_I(B\rightarrow j)
\label{standard}\ee

\noindent where $i$ denotes the $D\pi$ channel and $j$ any
hadronic channel, with the same quantum numbers as $D\pi$, the $B$
mesons can decay into. $H^{(2)}_W$ is the second order weak
hamiltonian and $\bar{A}_I$ are the bare transition amplitudes.

$H^{(2)}_W$ contains an isospin zero and an isospin one part. The
$\bar{A}_I$'s are directly related to the reduced matrix elements
of these specific isospin components of $H^{(2)}_W$. In the
absence of CP violation, the $\bar{A}_I$ are thus   relatively
real in any theory with hadronic asymptotic states.

From Eq.(\ref{standard}) we can now give a precise meaning to what
some of us have called the {\em quasi-elastic approximation}. It
is defined by the following equations:

\begin{equation}
S^{\frac{1}{2}}_{ii} = \sigma_i e^{i\delta_I}
\end{equation}

\noindent and

\begin{equation}
\sum_{j\neq i} S^{\frac{1}{2}}_{ij}\ \bar{A} (B\rightarrow
j)  =   0\ .
\end{equation}

In Eq.(4), $\sigma_i$ is a complex number. Its modulus is smaller
than one and  may depend on the isospin channel $I$, but we
specifically assume that its phase is isospin independent. On the
other hand,  $e^{i\delta_I}$ is the usual elastic $s$-wave phase
shift.

Substituting Eqs.(4) and (5) in Eq.(3) obviously implies that the
relative phase between $A_{\frac{1}{2}}$ and $A_{\frac{3}{2}}$ is
simply $e^{i(\delta_1-\delta_3)}$.

We strongly emphasize the fact that the quasi-elastic
approximation does by no means imply the absence of inelasticity:
it is \underline{not} assumed that $\sigma_i$ is of modulus one
nor that each $S^{\frac{1}{2}}_{ij} (i \neq j)$ vanishes. The
latter assumptions i.e. $|\sigma_i | = 1$ and
$S^{\frac{1}{2}}_{ij} (i \neq j) = 0$, correspond to the genuine
elastic limit which is of course physically absurd at the $B$ mass
\cite{dono}.

The quasi-elastic approximation as defined by Eqs.(4) and (5) does
not violate any basic principle. At ultra-high energies, for
example, all elastic amplitudes become purely imaginary and the
$\delta_I$'s tend to zero. In that limit Eqs.(4) and (5) implement
the physically sensible argument of Bjorken \cite{bjorken} that if
the $B$ meson were infinitely heavy there would simply be no time
for the final hadrons to rescatter. But no rescattering does not
mean no inelasticity!

The quasi-elastic approximation defines a {\em phenomenological
model} for final state interaction phases. The main virtues of
this model are that it puts so called "strong phases" where they
belong, namely in the hadronic world with its excellent isospin
symmetry and, more importantly, it allows for a simple calculation
of these phases in a Regge model of hadronic scattering. Other
models have of course been proposed for such strong phases in
$B$-decays, e.g. the random phase model \cite{suzuki} or the
coherent phase model \cite{Zen}.

As mentioned in the introduction, the quasi-elastic model was used
to analyze the decays $D \to \pi\pi, \pi K, K \overline K$ and the
results are in good agreement with the data. To our knowledge no
other model has met with similar successes.

We do assume that final state interactions are hadronic effects.
The implementation of this idea in the phenomenologically
well-defined quasi-elastic approximation works well for $D$
decays. We expect this phenomenology to be successful in $B$
decays as well.

\vspace*{5mm}

\par\noindent
{\Large \bf 3. A Regge model for $D\pi$ scattering}\vspace{0.5cm}

From Eqs.(3) and (5) it follows that $\delta_{D\pi} =
\delta_3-\delta_1$. To calculate this phase  we now use a simple
Regge model for elastic $D\pi$ scattering. Our notation and
parametrization will be the same as for the elastic $K\pi$
scattering treated in Ref. \cite{gerard1} where a more detailed
discussion can be found.

In the $t-$channel, $D\bar{D}\rightarrow \pi\pi$, the leading
Regge trajectories are the Pomeron (P) and the exchange degenerate
$\rho - f_0$ trajectory $(\rho)$. In the $u-$channel, the relevant
Regge trajectory would be that of the $D^\ast$ but since it lies
so much lower than the $\rho$ trajectory, it can safely be
neglected.

Following step by step the procedure outlined in Ref.
\cite{gerard1}, one obtains for the $l=0$ partial wave amplitudes

\ba a_{\frac{1}{2}}(s)\ &=&\ \frac{i}{\sqrt{6}}\frac{\beta_P
(0)}{b_P} s + \frac{1}{2} \frac{\bar{\beta}_{\rho}
(0)}{\sqrt{\pi}} \frac{1}{\ln s} s^{\frac{1}{2}} +
\frac{3i}{2\sqrt{\pi}} \bar{\beta}_{\rho}(0) \frac{\ln s +
i\pi}{(\ln s)^2 + \pi^2} s^{\frac{1}{2}}\nonumber\\
  a_{\frac{3}{2}}(s)\ &=&\ \frac{i}{\sqrt{6}}\frac{\beta_P
(0)}{b_P} s - \frac{\bar{\beta}_{\rho}(0)}{\sqrt{\pi}}\frac{1}{\ln
s} s^{\frac{1}{2}} \ea

\noindent from which the $\delta_I$ are easily computed. They
depend on one single parameter

\be x_{D\pi}\ =\ \frac{\sqrt{\pi} \beta_P (0)}{b_P
\bar{\beta}_{\rho} (0)} \label{parameter1} \ee

\noindent where $b_P$ is the slope of the Pomeron residue
function, while the couplings are $\beta_P(0) = g_{PD\bar{D}}\
g_{P\pi \pi}$ and $\bar{\beta}_{\rho}(0) = g_{\rho D\bar{D}}\
g_{\rho \pi \pi}$.

In Fig. 1 we plot $\delta_{D\pi}$ as a function of $x_{D\pi}$. In
the unrealistic SU(4) limit \cite{zheng} for $b_P$, $\beta_P$ and
$\bar{\beta}_{\rho}$  we would have $x_{D\pi} = x_{K\pi}$ close to
one \cite{gerard1} but, of course, we expect SU(4) to be badly
broken. In Eq.(1), the central value  of the CLEO data clearly
suggests $x_{D\pi}<1$.

\vspace*{5mm}

\par\noindent

{\Large \bf 4. The rescattering phase in $B\rightarrow DK$ decays}

\vspace{0.5cm}

We now consider the decays $B\rightarrow DK$.
There are two isospin amplitudes $A_0$ and $A_1$ and

\ba A(B^- \rightarrow D^0 K^-)\ &=& \ \sqrt{2}A_1\nonumber \\
A(\bar{B}^0 \rightarrow D^+ K^-)\ &=& \ +
\frac{1}{\sqrt{2}}\ A_0\ + \frac{1}{\sqrt{2}}\ A_1 \nonumber\\
A(\bar{B}^0 \rightarrow D^0 \bar{K}^0)\ &=& \
  - \frac{1}{\sqrt{2}}\ A_0  + \frac{1}{\sqrt{2}}\
A_1\ . \ea

Following the procedure given in Ref. \cite{gerard2} for
$K\overline K$ scattering, the $l=0$ partial wave amplitudes are
given by

\ba \tilde{a}_0(s)\ &=&\ \frac{i}{2}\
\frac{\tilde{\beta}_P(0)}{\tilde{b}_P} s \ +\
\frac{4i\tilde{\bar\beta_\rho}(0)}{\sqrt{\pi}}\ \frac{(\ln s)+i
\pi}{(\ln s)^2 + \pi^2}\ s^{\frac{1}{2}}\nonumber \\
\tilde{a}_1(s)\ &=&\ \frac{i}{2}\
\frac{\tilde{\beta}_P(0)}{\tilde{b}_P}\ s \ea

\noindent and the relevant parameter to determine the rescattering
phase $\delta_{DK} \equiv \delta_1 - \delta_0$ is now

\be \ x_{DK}\ =\ \frac{\sqrt{\pi}\
}{\widetilde{b}_P}\frac{\widetilde{\beta}_P(0)}{\widetilde{\bar
\beta}_{\rho}(0)} \ee where the couplings are $\tilde \beta_P (0)$
= $g_{PD\overline D}$ $g_{PK\overline K}$ and $\widetilde{\bar
\beta}_{\rho} (0)$ = $g_{\rho D \overline D}$ $g_{\rho K \overline
K}$.

From Eqs.(7) and (10) and using the data given in Ref.
\cite{gerard2}, we obtain

\begin{equation}
{x_{DK}\over x_{D\pi}} = {g_{PK\overline K} g_{\rho\pi\pi}\over
g_{P\pi\pi} g_{\rho K\overline K}} = 1.6 \pm 0.3
\end{equation}

\noindent in the SU(3) limit $b_P = \tilde b_P$. In fact, Eq.(11)
is compatible with a pure SU(3) estimate

\begin{equation}
\left({x_{DK}\over x_{D\pi}} \right)^{SU(3)} = 4/3.
\end{equation}

In Fig. 1 we also plot $\delta_{DK}$ as a function of $x_{D\pi}$.
If, for the sake of argument we take $0.2<x_{D\pi}<0.5$, then
$\delta_{DK}$ is predicted to be in the range

\begin{equation}
14^\circ < - \delta_{DK} < 24^\circ
\end{equation}

\noindent in the SU(3) limit defined by Eq.(12).

Clearly the model is compatible with the present data.

\vspace*{5mm}

\par\noindent
{\Large \bf 5. Conclusion} \vspace{0.5cm}

In this note we have derived  a simple relation between the final
state interaction phases for $B\to D\pi$ and $B \to DK$ decays in
the quasi-elastic approximation. Better data will allow for a
significant test of the point of view that final state interaction
phases are due to coherent hadronic effects.

In the Cabibbo-favored $B\to D\pi$ and Cabibbo-suppressed $B\to
DK$ decays, the dominant underlying quark diagram seems to be
\cite{belle2} particularly simple (tree-level approximation) and,
in fact, the final state interaction phases are not particularly
interesting per se \cite{Flei}. The situation is of course quite
different in $B\to K\pi, \pi\pi$ or $K\overline K$ decays where
the quark diagrams are more complicated (in particular with
one-loop penguin diagrams involved) and the physics much more
interesting. Final state interaction phases are then relevant not
only in an amplitude analysis but also in various CP violating
asymmetries. The quasi-elastic model predictions for the pattern
of $B\to K\pi$ direct CP-asymmetries were already  discussed
elsewhere \cite{gerard4}.

\vspace*{5mm}

\par\noindent
{\bf Acknowledgments}

\vspace*{5mm}

The author (G.C.) acknowledges the financial support from CONACyT
(M\'exico) under contracts 32429-E and 35792-E.  This work was supported by the
Federal Office for Scientific, Technical and Cultural Affairs through the
Interuniversity Attraction Pole P5/27.

\newpage

\begin{center}
\begin{figure}[h]
\centerline{\epsfig{file=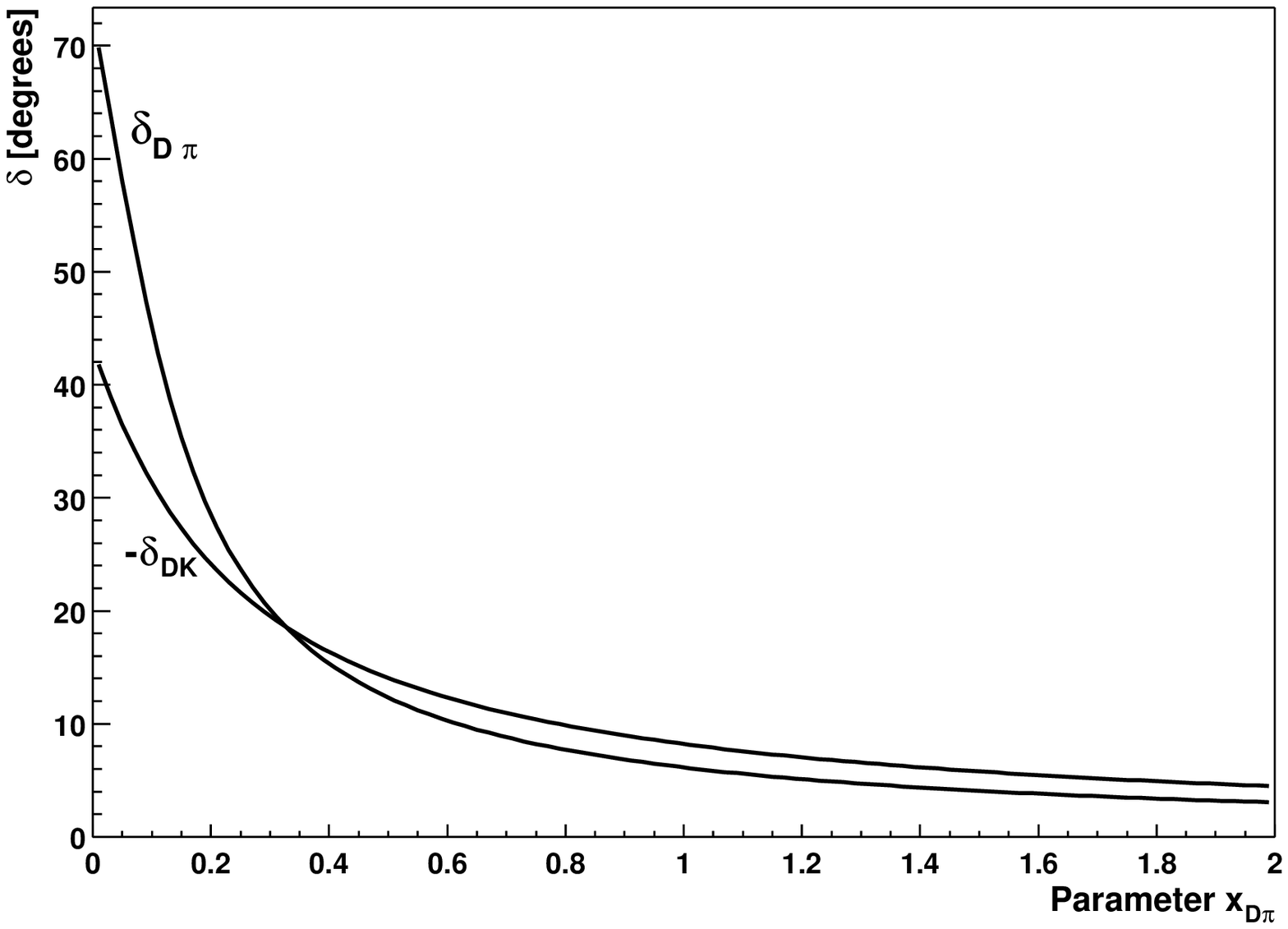,angle=0,width=12cm}}
\vspace{-0.5cm}\caption{Final state interaction phases in $B \to
D\pi$ and $B\to DK$ as a function of the Regge variable
$x_{D\pi}$ defined in Eq.(7).}
\end{figure}
\end{center}

\end{document}